\documentclass[10pt,twocolumn]{article}
\pdfoutput=1

\usepackage{etoolbox}
\patchcmd{\thebibliography}{\leftmargin\labelwidth}
    {\itemsep 0pt \leftmargin\labelwidth}{}{}

\usepackage[paper=a4paper,dvips,top=1.5cm,left=1.5cm,right=1.5cm,
    foot=1cm,bottom=1.5cm]{geometry}
    
\usepackage{graphicx}
\usepackage{mathptmx}
\usepackage{braket}
\usepackage{multicol}
\usepackage{lscape}
\usepackage{subfigure}
\usepackage{amssymb}
\usepackage{amsthm}
\usepackage{flushend}
\usepackage{hyperref}

\title{Composite Centrality:\\ A Natural Scale for Complex Evolving Networks}

\author{
        \textsc{Andreas Joseph}\thanks{\href{mailto:andreas.ch.joseph@gmail.com}{\texttt{andreas.ch.joseph@gmail.com}}}
            \qquad
        \textsc{Guanrong Chen}
        \mbox{}\\ %
         Centre for Chaos and Complex Networks, Department of Electronic Engineering \\ 
         City University of Hong Kong, 83 Tat Chee Avenue, Kowloon, Hong Kong SAR, P. R. China
}

\begin{document}

\begin{multicols}{1}
\maketitle
\end{multicols}

\begin{abstract}
\textbf{We derive a composite centrality measure for general weighted and directed complex networks,
based on measure standardisation and invariant statistical inheritance schemes. Different schemes generate different
intermediate abstract measures providing additional information, while the composite centrality measure tends to
the standard normal distribution. This offers a unified scale to measure node and edge centralities for complex 
evolving networks under a uniform framework. Considering two real-world cases of the world trade web and the 
world migration web, both during a time span of 40 years, we propose a standard set-up to demonstrate its 
remarkable normative power and accuracy. We illustrate the applicability of the proposed framework for large and 
arbitrary complex systems, as well as its limitations, through extensive numerical simulations.}\\[.2cm]

\end{abstract}
%

%
%
\section{Introduction}
\label{sec:I}
Starting with the work of Watts, Strogatz, Barab\'{a}si, Albert and Newman \cite{SW1,SF,SW2}, the investigation of complex networks has attracted an inflationary amount of attention from numerous research fields due to their ubiquity in the real world \cite{NW_book1,NW_book2,NW_book3}. One of the  fascinations lies in some elegant and efficient descriptions of very different complex systems under the general framework of modern graph theory \cite{GT_book1,GT_book2} pioneered by Erd\"{o}s \cite{RG}. \\
As the awareness of several common dynamic characters of many real-world networks rises, ever more effort has been devoted to understanding the temporal evolution of complex networks. Prominent examples of evolving networks are the internet and social networks \cite{NW_book2}, transportation networks \cite{trans_evo}, the world trade web \cite{WTW2,WTW3} and most recently climate networks \cite{climate_evo}. Any of such networks is generated by and/or hosts an underlying flow between its nodes, such as information, contacts, goods or diseases. Considering the description and analysis of evolving network structures, most efforts have been made regarding network modelling \cite{enwm_1,enwm_2,enwm_3,enwm_4,enwm_5}, while the development of sophisticated analysis tools and methodologies has seen less progress. This yet will be the topic of this work.\\
The characterisation and classification of general complex systems and especially complex evolving networks pose three major challenges:
\begin{itemize}
\item \textit{Uniformity:} There is a large variety of network measures stretching over wide numerical ranges, but there is no standardised procedure today to consistently consider several measures simultaneously. 
\item \textit{Variability:} Observed over time, many complex networks show growth (change of the number of nodes) and evolution (change of the topology). Network measures often depend explicitly on these quantities, which complicates a coherent temporal analysis.
\item \textit{Comparability:} There is no unified scale on which one can compare results originating from different networks.
\end{itemize} 
Furthermore, \cite{network_flow} and \cite{graph_theoretic_centrality} classified the underlying flows and the corresponding 
graph measures in terms of their physical or graph-theoretic properties, respectively. In a given situation, the exact details of the underlying flows might not be well-understood, and a multi-dimensional analysis of graph measures allowing for simultaneous evaluations is desirable. \\ 
Motivated by this and the above-stated general problems, we propose here a new centrality framework, called \textit{composite centrality} (CC). Generally, the notion of centrality can be understood as a measures quantifying the participation of a node (or any other component) in the underlying flow structure of a network \cite{network_flow}. This will also be the point of view we adopt in this work. The idea behind the CC-framework is that one first defines a set of characteristics of interest, and then chooses appropriate network (centrality) measures. The major complications when considering multiple network measures are different (often arbitrary) numerical scales and variously shaped distributions such as distributions with and without heavy tails. We therefore implement a \textit{standardisation procedure} involving a non-linear transformation and statistical normalisation. Relying on statistical methods, uniformity and variability are accounted for. Standardised measures can be  combined using \textit{invariant inheritance schemes} to form new standardised measures carrying abstract physical meanings, which we call \textit{composite centrality}. It turns out that final CC-scores for different set-ups and different networks are well-approximated by the standard normal distribution with a zero mean and a unit variance. This is what we call a universal scale to compare scores for different set-ups and even different networks across time (comparability).\\
This paper is structured as follows. In \hyperref[sec:II]{Section\ 2}, we give a short introduction to the relevant terminology from graph theory and propose a recipe for (graph) measure standardisation. In \hyperref[sec:III]{Section\ 3}, we present the CC-framework and introduce a specific standard framework. We demonstrate the working of the proposed set-up by considering two cases of the world trade web and the world migration web, both during a time span of 40 years. Furthermore, a graphical tool, which we call the \textit{network genetic fingerprint}, is introduced. It allows for efficient analysis and monitoring of composite centrality scores. In \hyperref[sec:IV]{Section\ 4}, we discuss the validity and limitations of the proposed framework through of large-scale simulations. We finally conclude the study in \hyperref[sec:V]{Section\ 5}.
\section{Preliminaries}
\label{sec:II}
\subsection{Graph Theory}
\label{sec:II-A}
In this section, we give a short introduction to the parts of graph theory that are needed in the following. Explicit formulas will be given only if deemed necessary. For a more detailed introduction, we refer to \cite{NW_book1,NW_book2,NW_book3,GT_book1}. \\
Graph theory provides a general mathematical framework to represent and quantify complex networks and their properties. A weighted and directed network can be represented by a graph $G\,=\,\left(V,\,E\right)$, where $V\,=\,\lbrace v_1,\, \dots,\, v_N \rbrace$ is the set of $N\geq 2$ nodes (vertices) in the graph. $E\left(w_{ij}>0|\,i,j\in\lbrace 1,\,\dots,\,N\rbrace\right)$ is the set of weighted edges from node $v_i$ to node $v_j$, with $N_e\,=\,ord(E)$ denoting the number of edges irrespective of their weights. The whole graph can be represented by a real weight matrix, $W\,=\,[w_{ij}]\,\in\,\mathbb{R}^{N\times N}$ ($w_{ij}\,\neq\,w_{ji}$, in general). We do not allow for self-loops here, i.e.\ $w_{ii}\,=\,0$ for all $i\,\in\,\lbrace 1,\,\dots,\,N\rbrace$. The  $i^{\mathrm{th}}$ row or column represents the out- or in-strength distribution of node $i$, respectively. The \textit{total strength} of a node $i$, denoted by $s_i$,  is the sum of in-strengths $s^{\mathrm{in}}_i$ and out-strengths $s^{\mathrm{out}}_i$ of that node. It represents a generalisation of the degree centrality (number of adjacent edges) in an undirected and unweighted graph. The degree of a node can be obtained from the underlying simple (non-weighted, non-directed, no self-loops) graph through its adjacency matrix $A\,=\,[a_{ij}]\equiv[a_{ji}]\,\in\,\lbrace 0,\,1 \rbrace$, where $a_{ij}=1$ if there is an edge between node $i$ and node $j$, but $a_{ij}=0$ otherwise. Likewise, since self-loops are not allowed, one has $a_{ii}=0$ for all $i\in\{1,\,\dots,\,N\}$. The degree of node $i$ is given by the $i^{\mathrm{th}}$ row or column sum of $A$. Strength and degree of a node can be interpreted as two measures for local connectivity either considering weighted or unweighted graphs, respectively. Here and later, we refer to measures over weighted networks as being of  \textit{quantitative} nature and measures over unweighted networks as being \textit{qualitative}. The difference between both levels of complexity is summarised under the notion of (edge) \textit{texture}. It is said that there is a connection between any two nodes $i$ and $j$ in $G$ if there exists a directed path $p_{ij}$ from $i$ to $j$ ($p_{ij}\neq\,p_{ji}$, in general). A directed graph is said to be \textit{strongly-connected} if there exists a directed path between any two nodes. This means that the weight matrix $W$ and the adjacency matrix $A$ are both irreducible. A measure for the connectivity of a graph on the global scale is the \textit{edge density} $\rho_e\,=\,N_e/\left(N^2-N\right)$ (number of actual edges divided by number of possible edges), while on a local scale the embedding of a node can be expressed via its \textit{clustering coefficient}. On the adjacency level, the clustering coefficient of a node is defined as the number of actual connections within its neighbours over the number of possible connections among them. A further important measure to quantify the participation of a node $i$ into the path-structure of a network indicating overall connectivity is the \textit{average shortest path length} per other node $l_i=\braket{l_{ij}}_i$ (or \textit{farness}), i.e.\ the average number of steps (over unweighted edges) which it takes to get from node $i$ to any other node $j$. A generalisation to weighted edges is straightforward, once one relates edge weight to distance. Note that in a directed graph the shortest path between two nodes is generally not symmetric, i.e.\ $l_{ij}\neq\,l_{ji}$. The \textit{diameter} $\o$ of a graph is defined as the maximal shortest path between any two nodes. The \textit{maximal flow} $f_{ij}$ between two nodes $i$ and $j$ is the maximal capacity that can be transported parallelly from node $i$ to node $j$ via the whole graph for networks in which the edge weight can be interpreted as representing some form of capacity \cite{NW_book1}, e.g.\ bandwidth for electronic data transmission (again  $f_{ij}\neq\,f_{ji}$, in general). \\
\textit{Graph asymmetry} is a measure for the difference between $w_{ij}$ and $w_{ji}$ on a global scale, i.e.\ for the overall weight balance. We define it as 
\begin{equation}
\label{eq:asy}
A_D\,=\,\frac{||W-W^T||_F}{2\,||W||_F}\,\in\,[0,1]\,,
\end{equation}
where $||\cdot ||_F$ denotes the Frobenius norm of a matrix. The \textit{algebraic connectivity} $\lambda_1$ of the normalised Laplacian $L_N\,=\,D^{-\frac{1}{2}}\left(D-W\right)D^{-\frac{1}{2}}$, where $D\,=\,diag(s_1,\,\dots,\,s_N)$ is a diagonal matrix consisting of the nodes' strengths, is the smallest non-zero eigenvalue, while the Laplacian always has a single zero eigenvalue for the case of only one single connected component. It is a measure for the robustness of a graph against node removal (failure) \cite{NW_book1,lambda1}. A further measure is \textit{assortativity} \cite{NW_book1,NW_book2}, $As\,\in\,[-1,1]$, which describes the overall homogeneity of connections indicating if weak/strong nodes are preferentially coupled to other weak/\\strong nodes (or vice versa), resulting in a positive (or negative) $As$ value. \textit{Eigenvector centrality} of a node $i$ is defined as the $i^{\mathrm{th}}$ entry of the eigenvector corresponding to the largest eigenvalue of the underlying graph's adjacency matrix. It measures how well a node is connected to the whole graph or to other well-connected (high-scoring) nodes recursively \cite{NW_book1}. \\
\subsection{Measure Standardisation}
\label{sec:II-B}
Different node measures (e.g. centrality measures) generally span wide and different numerical ranges, show different levels of variation and exhibit variously shaped distributions, which makes them difficult to compare\footnote{Here and throughout, we only consider node measures. A generalisation to edge measures is straightforward.}. For instance, it is difficult to simultaneously treat measures with exponentially decaying (e.g.\ Gaussian) and a heavy (e.g.\ power-law) tails, which actually poses a standard problem for analysing evolving networks. \\
\indent We address these issues by defining a standardisation procedure for non-trivial positive-valued and (approximately) uni-modal measures. The main idea behind this transformation is to \textit{harmonise} the first, second and third moments of any given measure distribution, which are addressed in three or four steps:
\begin{enumerate}
\item \textbf{Skewness:} 
   \begin{itemize}
   \item Rescale to a mean of \textit{one}.
   \item Perform a Box-Cox transformation.
   \item Accept the Box-Cox-transformed measure, only if the skewness could be reduced.
   \end{itemize}
\item \textbf{Mean:} Shift to a \textit{zero} mean.
\item \textbf{Variance:} Divide all values by the sample standard deviation.
\item \textbf{Order (if necessary):} Mirror all values with respect to the origin, such that all measures follow the rule of \textit{bigger-is-better}.
\end{enumerate}
Step 1 addresses the issue of variously shaped measure distributions, in the sense that it aims at bringing a sample's skewness to zero via a bijective and non-linear Box-Cox power transformation \cite{box_cox_trafo,box_cox_rev}. It is defined as 
\begin{equation}
\label{eq:bct}
\tilde{x}\,\equiv\,\left\{ 
  \begin{array}{l l}
    \frac{x^{\lambda}-1}{\lambda}& \qquad \mbox{if $\lambda\neq 0$}\\
    \ln x & \qquad \mbox{if $\lambda = 0\,,$}
  \end{array} \right.
\end{equation}
while the real parameter $\lambda$ maximises the log-likelihood function
\begin{equation}
\label{eq:ll}
\log\mbox{-}L \,=\, \left(\lambda-1\right)\sum_i \ln\, x_i \,-\, \frac{N}{2}\,\ln\, \sum_i \frac{\left(\tilde{x}_i-\braket{x}\right)^2}{N}\,.
\end{equation}
The rescaling to one is done to enhance comparability between transformed measures, since one is the only point projected to zero, irrespective of the exponent $\lambda$. Furthermore, transformation \ref{eq:bct} is approximately linear around one, such that only a distribution's tails are affected by the non-linearity of the distribution. In the presence of negative values, one might shift the whole distribution slightly up into the positive realm, which is expected to give similar results as a two-parameter Box-Cox transformation \cite{box_cox_rev}.\\
Steps 2 and 3 are what we call statistical normalisation. Transformation \ref{eq:bct} results in a distribution which is approximately normal in general, but does not fix the mean or the standard deviation of the transformed values. By dividing through the sample standard deviation, one takes into account the problem of different units, which are often arbitrary. Given a representative sample (see below), standardised measures are independent of the sample size, which makes them especially suited for the investigation of evolving networks with a varying number of nodes and edges.\\
Step 4 is applied to impose the same numerical ordering on all measures. We presume that all measures should follow the principle of \textit{bigger-is-better}, i.e.\ larger numerical values are higher ranked. A simple ordering problem is given for example when one wants to compare nodes' degrees and farness (abbreviated as ASPL). For the former measure a larger value is generally higher ranked, while for the latter the opposite is the case.\\
In summary, measures, which have been standardised according to the above recipe, have the following characteristics: zero mean, unit variance, zero skewness and a Gaussian shape. We note that the latter two properties are expected to be only approximately accurate. The Gaussianity of standardised measures will be tested extensively later (see sections \ref{sec:III-B} and \ref{sec:IV}).\\ 
One very interesting feature of standardised measures is that such variational measures (i.e. expressed in terms of standard deviations) indicate how good/bad a node's score is in comparison to all other nodes, which might be of interest when investigating the evolution of  scores over time, independent of any varying absolute scales, e.g.\ inflating prices. Due to their shared properties scores for standardised measures can be directly compared to each other, even between different networks. 
Note furthermore that the presented standardisation recipe is reversible, i.e.\ one can always go back to the original measures. Besides, for the here-presented applications of uniformly comparable network measures, standardised measures can be used for well-defined correlation and regression analyses - general methodologies facing similar problems as identified for the study of complex networks.
\section{The CC-Framework}
\label{sec:III}
\subsection{Composite Centrality}
\label{sec:III-A}
By only using standardised measures (SM), it is straightforward to construct arbitrary composite measures. Having two node measures, $A$ and $B$, for which $\bar{A}$ and $\bar{B}$ denote the corresponding SM, the composite centrality measure is defined by the statistically normalised measure of the sum of the two measures via value-by-value addition. One has
\begin{equation}
\label{eq:cc}
C_{\mathrm{comp}}\left(A,B\right)\,\equiv\,\frac{\bar{A}+\bar{B}}{\sigma_s\left(\bar{A}+\bar{B}\right)}\,,
\end{equation}
where $\sigma_{s}(\cdot)$ stands for the sample standard deviation. $C_{\mathrm{comp}}$ is a new SM carrying an abstract physical meaning depending on the original measures. From here on, standardised measures are denoted with a bar. Having a set of $n$ measures, $M \ni M_i$, $i\in \lbrace 1,\,\dots,\,n\rbrace$, a generalisation of (\ref{eq:cc}) takes the form
\begin{equation}
\label{eq:ccc}
C_{\mathrm{comp}}\left(M\right)\,=\,\frac{\sum_{i=1}^{n} \bar{M}_i}{\sigma_s\left(\sum_{i=1}^{n} \bar{M}_i\right)}\,.
\end{equation}
%
It is independent of the order which the $\bar{M}_i$ are combined, up to negligible statistical fluctuations. We may derive one further result from the form of $C_{\mathrm{comp}}$. Provided that the assumptions of the Lyapunov theorem \cite{Lyap_thm} hold\footnote{This may not always be the case; see \hyperref[sec:IV]{Section\ 4}.}, the central limit theorem (CLT) from statistics \cite{Lyap_thm,prob_book1} can be extended to include \textit{non-identically} distributed random variables. We may then state that for a set of $n$ independent random variables, which are represented by the SM $\bar{M}_i$, the sampled random variable $\bar{M}_{\mathrm{comp}}$ tends to a standard normal distribution with zero mean and unit variance as the sample size (i.e.\ the number of measures) increases. This sets a \textit{unified scale} for composite centrality measures, because the limiting distribution is \textit{parameter-free}, thus rendering node composite centrality scores comparable over time and for different networks (data sets).\\
\subsection{A Standard Framework: Direction, Range and Texture}
\label{sec:III-B}
\begin{figure}[!t]
\centering
\includegraphics[width=3.in]{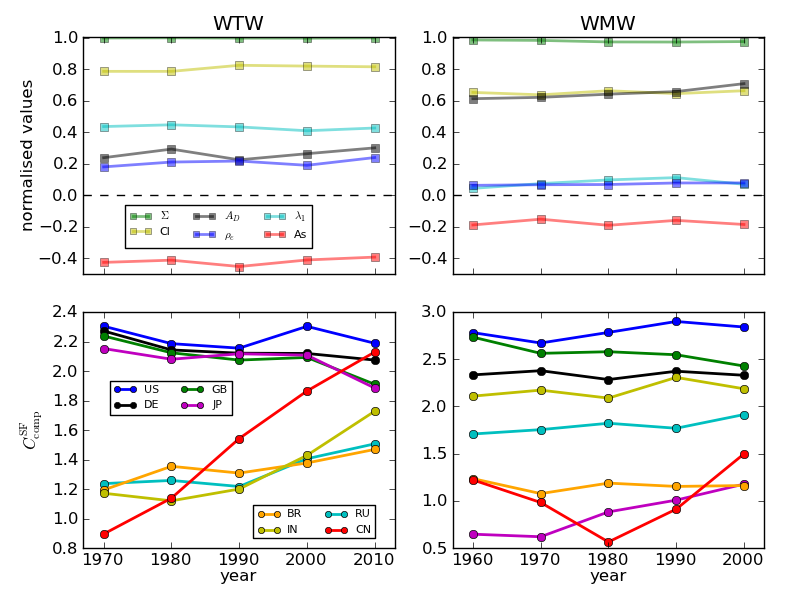}
\caption{(colour online) Upper row: Normalised measures: coverage $\Sigma$ (fraction of edge data included in LSCTG),  clustering coefficient of the underlying simple graph $Cl$, algebraic connectivity $\lambda_1$, graph asymmetry $A_D$, edge density $\rho_e$, assortativity $As$. Bottom row: Evolution of $C_{\mathrm{comp}}$ for selected countries (see text); WTW (left) and WMW (right). Country codes are given using the ISO 3166-1 2-digit standard.}
\label{fig:props}
\end{figure}

The formulation of the CC-framework has so far been very general. In this section, we specify a certain set-up, which we propose as
a standard framework (SF) for the investigation of weighted and directed networks. We demonstrate the working of the SF by analysing data from the world trade web (WTW) \cite{comtrade} and the world migration web (WMW) \cite{worldbank}, which can both be characterised as socio-economic networks. In both networks, nodes are countries and territories. In the WTW, edges are aggregated directed trade flows valued in USD reported during one calendar year, while in the WMW, edges are directed migrant flows reported since the last census. Data have been raised every year and every ten years, respectively. We consider observation periods of 40 years sliced into 10-year intervals for both networks\footnote{Basically, the maximal time span for which data are available.}; WTW: 1970-2010, WMW: 1960-2000. Despite the fact that both networks are very different in nature and also in structure, it turns out that the SF is well-suited to analysing both networks, regarding their properties as well as their temporal evolution.\\
In the SF, we are interested in a node's centrality based on the physical criteria of \textit{direction} (in- and out-bounded connectivity, D: IN/OUT), \textit{range} (long- and short-range connectivity, R: LO/SH) and \textit{texture} (qualitative and quantitative connectivity, S: QL/QN). To achieve a high centrality score, a node must score well regarding all criteria. Since all three criteria are binarily divided, we need a total of $2^3=8$ measures to characterise them. Using a specification of the form D-R-T, a possible choice of measure $M_{\mathrm{SF}}$ is given in \hyperref[tab:SF measures]{Tab.\ 1}. Note that we focus on radial \cite{graph_theoretic_centrality} measure, where the node of interest sits at the endpoints of a certain path. Those measures can be used to describe the influence a node has according to the given criteria. Another class of measures are medial \cite{graph_theoretic_centrality} measure, where the node interest sits in the middle of a certain path. Such measures can be used to evaluate the amount of control a certain node can exert, while the application of the CC-framework is straightforward thanks to its generality. \\
\begin{center}
\begin{table}[h]\renewcommand{\arraystretch}{1.5}
\begin{center}
\begin{tabular}{l|l|l}
D - R - T & description & symbol \\
\hline
\hline
IN-LO-QL & incoming ASPL & $l_{\mathrm{in}}$ \\
IN-LO-QN & incoming max. flow & $f_{\mathrm{in}}$ \\ 
IN-SH-QL & in-degree  & $d_{\mathrm{in}}$ \\
IN-SH-QN & in-strength  & $s_{\mathrm{in}}$ \\
OUT-LO-QL &  outgoing ASPL  & $l_{\mathrm{out}}$ \\
OUT-LO-QN &  outgoing max. flow & $f_{\mathrm{out}}$  \\  
OUT-SH-QL &  out-degree  & $d_{\mathrm{out}}$ \\
OUT-SH-QN &  out-strength  & $s_{\mathrm{out}}$\\
\end{tabular}
\end{center}
\caption{Radial SF measures based on the properties of direction (D), range (R) and texture (T).}
\label{tab:SF measures}
\end{table}
\end{center}
For both networks, we consider the largest strongly-connected component of a threshold graph (LSCTG), i.e.\ edges with a strength below a certain threshold $e_{\mathrm{th}}$ are removed. This has the advantage to reduce the relative errors in the data, and focuses the analysis on quantities of a certain minimal magnitude. The corresponding edge thresholds are set to $e_{\mathrm{th}}^{\mathrm{WTW}}=10^7\,$USD and $e_{\mathrm{th}}^{\mathrm{WMW}}=2000\,$migrants for the final year of observation, respectively. For other time instances, threshold values have been adjusted for world-GDP and -population growth, using the GDP-deflator \cite{un_deflator} and population growth rates \cite{un_pop_grow}, respectively. This has been done to allow for better comparability of analyses at different times and shows how network properties evolved as compared to those quantities. General properties of both networks and their evolution over time are shown in \hyperref[tab:num-vals]{Tab.\ 2} and in \hyperref[fig:props]{Fig.\ 1} (upper row). One can see that both evolving networks grow over time relative to world-GDP or -population. However, relative growth is weaker for the WMW than for the WTW. Graph asymmetry is considerable higher for the WMW (indicating preferred directions of migration), while negative assortativity is stronger for the WTW (indicating trade between ``unequal partners''). The differences between the two networks is summarised as follows: The WTW can be described as \textit{homogeneous} (high density and clustering, modest asymmetry, high algebraic connectivity and small diameter), while the WMW can be characterised as \textit{heterogeneous} (low density, modest clustering, high asymmetry, low algebraic connectivity, large diameter). \\
Having defined the set-up in terms of $M_{\mathrm{SF}}$, it is straightforward to standardise them using the recipe given in \hyperref[sec:II-B]{Section\ 2.2}. As pointed out, the distribution of $\bar{C}_{\mathrm{comp}}$ should be (approximately) standard normal, i.e.\ with mean $\mu=0$ and standard deviation $\sigma=1$ and a probability density of the form
\begin{equation}
\label{eq:logn}
p_{\mathrm{SN}}\left(x\right)\,=\,\frac{1}{\sqrt{2\pi}}\,\exp\left[-\,\frac{x^2}{2}\right],
\end{equation}
where $x$ is a random variable. This represents a universal centrality distribution in the sense that there is no free parameter involved, providing a unique scale for standardised (composite) centralities for evolving networks.\\
\begin{figure}[!t]
\centering
\includegraphics[width=3.in]{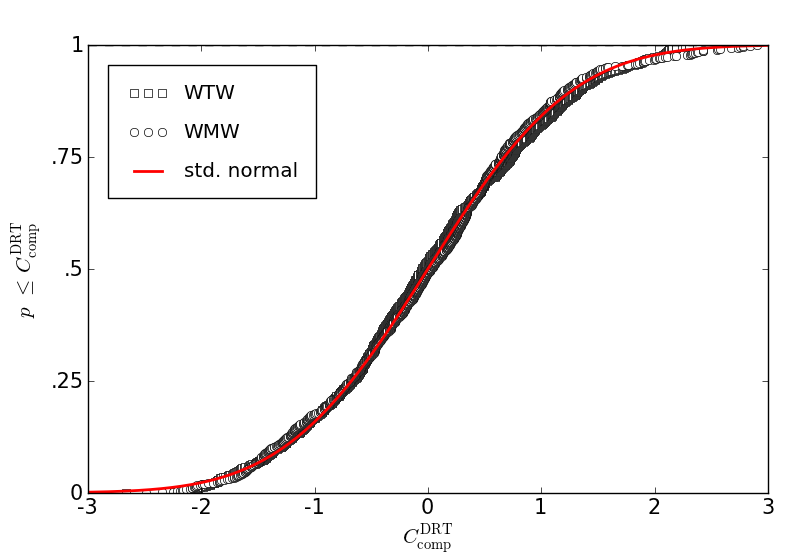}
\caption{(colour online) Cumulative distribution function (CDF) for composite centralities for the WTW (circles) and the WMW (squares). To demonstrate that data throughout different years (networks) can be treated uniformly, we combined scores for the corresponding 40-year observation periods of both networks. The red line shows the CDF of a standard normal distribution, which is seen to fit the data well. The $p$-values from the corresponding KS-tests are $.71$ and $.66$, respectively.}
\label{fig:fit}
\end{figure}
\begin{center}
\begin{table}[h!]\renewcommand{\arraystretch}{1.5}
\centering
\begin{tabular}{l|l}
distribution & parameters \\
\hline
\hline
uniform & $x_{\mathrm{inf}}=0\,,\,x_{\mathrm{max}}=1$ \\
normal & $\mu=10^5\,,\,\sigma=10^3$ \\
log-normal & $\mu=2\,,\,\sigma=2$ \\
exponential & $\mu=10^{-3}$ \\
Pareto & $x_{\mathrm{min}}=10^{2}\,,\,\alpha=3$\\
\multicolumn{2}{c}{}
\end{tabular}
\caption{Probability distributions contributing to $M_{\mathrm{arb}}$. In the right column, the numerical values for generic parameters are given which have been used to generate samples of size $N\,=\,10^2-10^4$.}
\label{tab:arb}
\end{table}
\end{center}
\begin{center}
\begin{table}[h]\renewcommand{\arraystretch}{1.5}
\centering
\begin{tabular}{l|rrrrr}
\multicolumn{6}{c}{WTW : $e_{\mathrm{th}}\,=\,10^7\,$USD}\\
\hline
\hline
year  & 1970 & 1980 & 1990 & 2000 & 2010 \\
\hline
$N$  & $181$ & $187$ & $179$ & $207$ & $204$ \\
$N_e$  & $2078$ & $4075$ & $4632$ & $6941$ & $9667$ \\
$\o$  & $4$ & $3$ & $3$ & $3$ & $3$ \\
$\hat{l}$  & $1.92$ & $1.86$ & $1.82$ & $1.88$ & $1.82$ \\
$\hat{f}/10^8$  & $0.14$ & $0.84$ & $1.40$ & $1.82$ & $5.44$ \\
$\hat{d}$  & $32.3$ & $39.1$ & $38.5$ & $39.0$ & $48.5$ \\
$\hat{s}/10^8$  & $0.09$ & $0.53$ & $1.00$ & $2.40$ & $3.42$ \\
\multicolumn{6}{c}{}
\end{tabular}
\begin{tabular}{l|rrrrr}
\multicolumn{6}{c}{WMW : $e_{\mathrm{th}}\,=\,2\cdot 10^3\,$migrants}\\
\hline
\hline
year  & 1960 & 1970 & 1980 & 1990 & 2000 \\
\hline
$N$  & $178$ & $179$ & $185$ & $182$ & $193$ \\
$N_e$  & $1957$ & $2107$ & $2292$ & $2544$ & $2880$ \\
$\o$  & $6$ & $6$ & $6$ & $6$ & $6$ \\
$\hat{l}$  & $2.56$ & $2.59$ & $2.50$ & $2.46$ & $2.42$ \\
$\hat{f}$  & $354$ & $440$ & $505$ & $717$ & $810$ \\
$\hat{d}$  & $11.0$ & $12.8$ & $12.4$ & $14.0$ & $14.9$ \\
$\hat{s}/10^4$  & $2.89$ & $3.24$ & $3.42$ & $4.16$ & $4.37$ \\
 \multicolumn{6}{c}{}
\end{tabular}
\caption{Nominal Numeric values for selected standard measures for WTW (upper table) and WMW (lower table): number of nodes $N$, number of edges $N_e$, diameter $\o$, ASPL $\hat{l}$, average max. flow per node $\hat{f}$, average degree $\hat{d}$, average strength per node $\hat{s}$.}
\label{tab:num-vals}
\end{table}
\end{center}
To demonstrate the applicability of the SF, we compute $\bar{C}^{\mathrm{SF}}_{\mathrm{comp}}$ for the WTW and the WMW, and investigate their evolution during a period of $40$ years. In \hyperref[sec:IV]{Section\ 4}, this analysis is extended to larger sets of synthetic data. \\
The extent to which $C^{\mathrm{SF}}_{\mathrm{comp}}$ can be described by the standard normal distribution (\ref{eq:logn}) is shown in \hyperref[fig:fit]{Fig.\ 2}, where data points from all years have been combined for better representation\footnote{This is possible since they are all expected to follow the same distribution.}. The red line shows the cumulative distribution function (CDF) of (\ref{eq:logn}), while the CDF of $\bar{C}^{\mathrm{SF}}_{\mathrm{comp}}$ for WTW and WMW are given by unfilled circles and squares, respectively. One can see that the Normal distribution fits the data well in both cases. To quantitatively test the standard normal hypothesis (SNH), we perform a \\goodness-of-fit (GoF) analysis by calculating $p$-values to $\epsilon<0.01$ precision using the Kolmogorov-Smirnov test (KS-test) \cite{KS_1}: The KS-test compares the maximal vertical distance (KS-statistic) between the empirical CDF and the hypothetical distribution's CDF to the corresponding distance for a set of synthetic samples. The $p$-value is the fraction of those samples where the empirical CDF is closer to the hypothetical CDF. Next, we define a decision rule (DR) for the acceptance or rejection of the SNH (null hypothesis). Following \cite{KS_1}, the null hypothesis is accepted for a $p$-value of $p>0.1$, potentially making a type-2 error.\\
The resulting $p$-values are $p^{\mathrm{WTW}}=.71$ and $p^{\mathrm{WMW}}=.66$. Hence, according to the above-defined decision rule, the SNH can be accepted. We conclude that it is indeed possible to describe node centralities and their evolution over time using the universal distribution (\ref{eq:logn}), providing a very useful tool for a uniform investigation of complex evolving networks.\\
To illustrate the evolution of $C^{\mathrm{SF}}_{\mathrm{comp}}$ and to demonstrate what one can learn from a composite centrality analysis, we consider \hyperref[fig:props]{Fig.\ 1} (lower low): The evolution of $C^{\mathrm{SF}}_{\mathrm{comp}}$ for a group of eight selected countries in the WTW (left) and the WMW (right) is shown, respectively. The four countries, United States (US), Germany (DE), Great Britain (GB) and Japan (JP), represent industrialised economies, while Brazil (BR), Russia (RU), India (IN) and China (CH) form the BRIC-block of developing economies. \\
Let us first consider the WTW. One can clearly see how the two groups converge after 1990. This convergence can essentially be interpreted as an illustration of globalisation. Note that the rise-and-fall of a score does not necessarily reflect an absolute change of some measure(s), but rather, how the score has changed in comparison to all other nodes. If we consider a network as a closed system, this may be all that matters. Furthermore, one may (carefully) extrapolate how these scores evolve into the future, what might be of great importance in considering global economic development and for tackling the challenges which might in turn arise. \\ 
A similar, but less pronounced behaviour is observed in the WMW. Observing the overlapping time window for both networks (1970-2000), one can see that the evolution in one network is not necessarily reflected in the other. Such comparisons are possible, since node centralities in both networks are measured using the same scale. Since both networks share a considerable fraction of nodes, this universality allows us to draw conclusions comparing node centralities in the WTW and in the WMW. For instance, the US is (still) the most central node to both networks. Regarding its much higher CC-value in the WMW, one concludes that the US is much more central to the WMW than it is to the WTW, from which, taken by itself, a multitude of implications and further questions may arise. In addition, the large economic development of China after 1990, which is well represented by the evolution of its CC-score in the WTW, is what we might call a \textit{truly Great Leap Forward}. This economic development is (so far) only weakly reflected in the WMW. An interesting question for future research is how a node's centrality in one network is related to its centrality in another network; and, if there are significant relations between centralities in different networks and how these relations come about. To answer such questions consistently, i.e.\ to be ably to do a thorough and effective analysis of such relations, we will present a novel graphical tool for the evaluation of the evolution of composite centrality scores in the \hyperref[sec:III-C]{next section}. \\
The above given examples demonstrate clearly how $C^{\mathrm{SF}}_{\mathrm{comp}}$ provides a very useful tool to evaluate and compare node centralities for different evolving networks. One future task will be to test the applicability of the proposed framework to a larger variety of complex networks, or even very general complex systems. This point will be further addressed in \hyperref[sec:IV-B]{Section\ 4.2}.
\subsection{Network Genetic Fingerprint}
\label{sec:III-C}
\begin{figure}[!t]
\centering
\includegraphics[width=3.in]{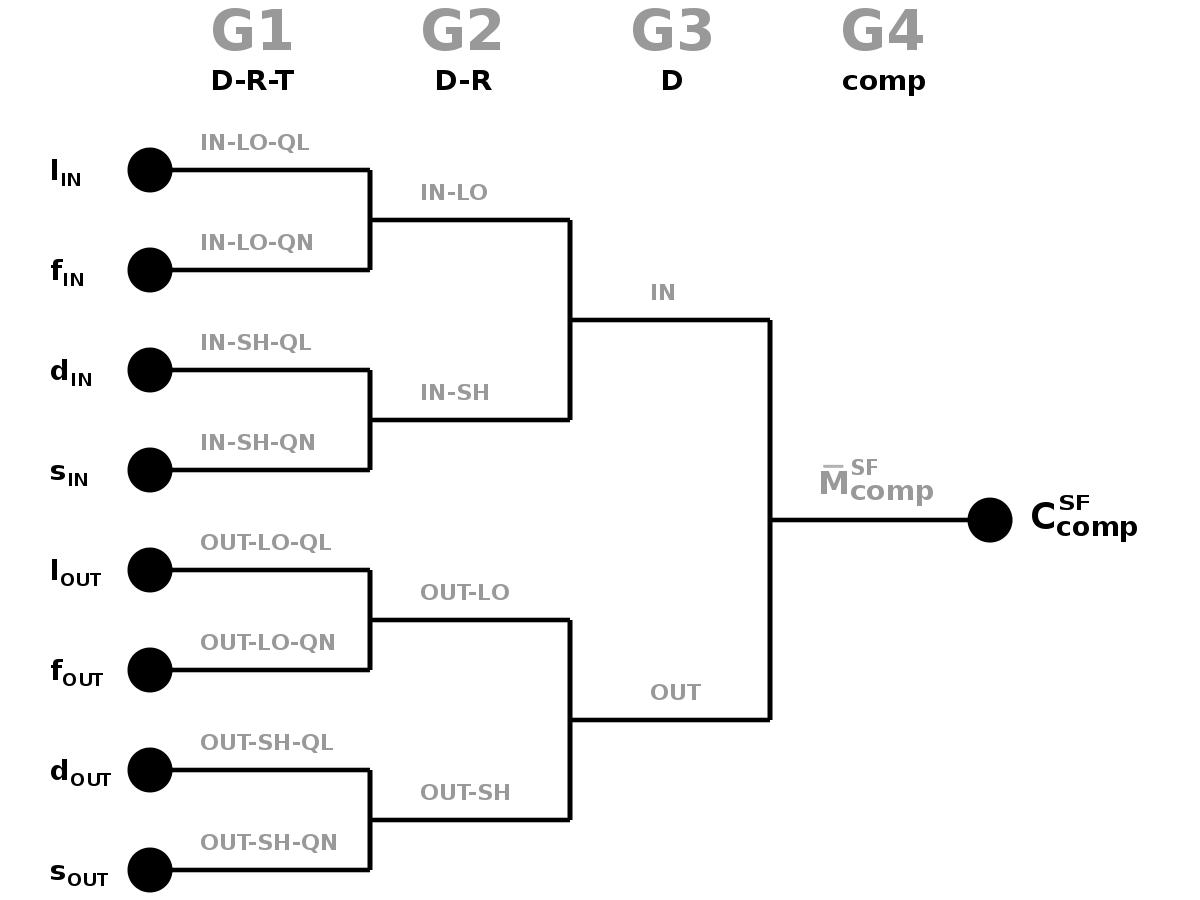}
\caption{Possible D-R-T inheritance scheme (see \hyperref[tab:SF measures]{Tab.\ 1}): First generation measures (G1) are successively combined to higher-generation measures (G2, G3, G4) using (\ref{eq:cc}), whereby intermediate measures provide additional information. $\bar{M}_{\mathrm{comp}}$, and hence $C_{\mathrm{comp}}$, are scheme-independent.}
\label{fig:gen}
\end{figure}

\begin{figure}[!t]
\centering
\includegraphics[width=3.in]{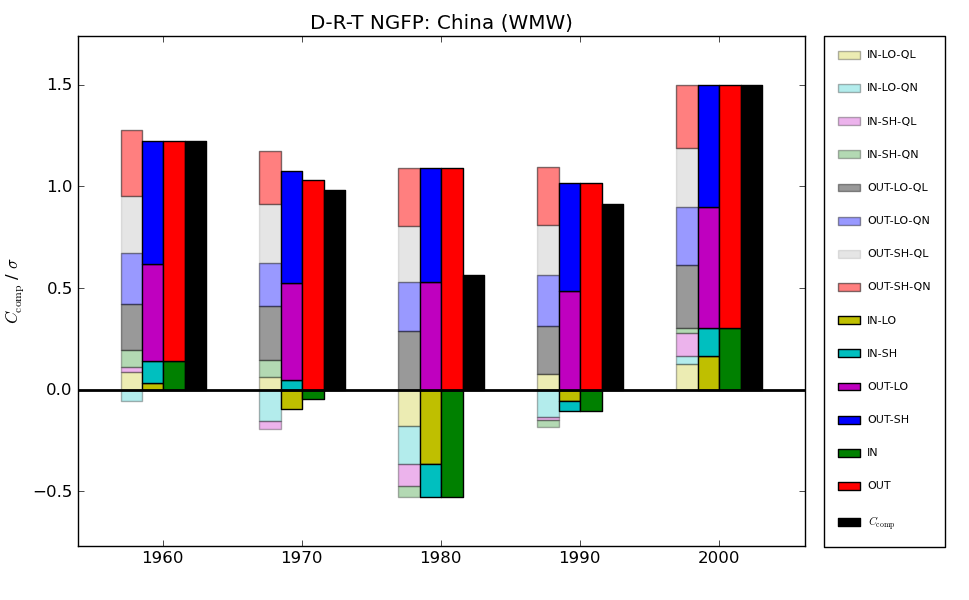}
\caption{(colour online) NGFP for China in the WMW for the years 1960-2000. For each year, from left to right, the relative bar heights stand for the G4-, G3-, G2- and G1-measures as described by the inheritance scheme shown in \hyperref[fig:gen]{Fig.\ 3}. Total heights (positive/negative) on the right (G1, G2, G3) sum up to heights on the left (G2, G3, G4), respectively.}
\label{fig:ngfp}
\end{figure}

Given a set of measures, $M$, the form of (\ref{eq:cc}) and (\ref{eq:ccc}) allows for the possibility to derive a variety of intermediate SM while computing $\bar{C}_{\mathrm{comp}}$, whereby the final value of $\bar{C}_{\mathrm{comp}}$ is independent of this ordering. A certain (invariant) ordering is called an \textit{inheritance scheme}. To make this point clear, we consider \hyperref[fig:gen]{ Fig.\ 3}, where we start with the $2^3=8$ measures from $M_{\mathrm{SF}}$ in the first generation (G1). In each subsequent step, two SM are combined according to (\ref{eq:cc}), generating sets of abstract higher generation measures, (G2, G3, G4), while the single measure in the forth generation (G4) equals $\bar{C}_{\mathrm{comp}}^{\mathrm{SF}}$ for all schemes.\\
The D-R-T \textit{inheritance scheme} for $M_{\mathrm{SF}}$ of \hyperref[fig:gen]{ Fig.\ 3} is defined as follows: Firstly, we combine T-measures (QL+QN having the same D-R-specification) to generate the second generation measures (G2) characterising abstract D-R-scores. Secondly, we combine R-measures (SH+LO having the same D-specifica-\\tion). This results in measures characterising the in- and out-bounded centrality of a node, while each score reflects now information originating from four single measures. Finally, we combine the remaining two D-measures to  obtain $\bar{M}^{\mathrm{SF}}_{\mathrm{comp}}$. In this way, one arrives at $C^{\mathrm{SF}}_{\mathrm{comp}}$ iteratively using (\ref{eq:cc}), while different schemes lead to the same final numerical values, up to minor statistical fluctuations. For the WTW and the WMW, these fluctuations of $C^{\mathrm{SF}}_{\mathrm{comp}}$ are of order $\mathcal{O}^{\mathrm{WTW}}\left(10^{-4}\right)$ and $\mathcal{O}^{\mathrm{WMW}}\left(10^{-3}\right)$, respectively. While G1- and G4-measures are the same in the case at hand, G2- and G3- measures of different inheritance schemes provide additional and complementary information regarding the individual role of nodes in a network and their evolution over time.\\
In this section, we present a graphical tool which enables one to easily monitor and analyse individual contributions to a node's composite centrality and evolution, when given a certain scheme. \hyperref[fig:ngfp]{Fig.\ 4} shows what we call the \textit{network genetic fingerprint}\\(NGFP) for China in the WMW using the D-R-T scheme. For each year, the bars stand for G4-, G3-, G2- and G1-SM from left to right, respectively. The relative height of each coloured region, measured from the zero expectation (black horizontal line), stands for the corresponding SM's magnitude in $\sigma_{\mathrm{comp}}$.\\
We remark that, for a G1-measure, this zero line, when transformed back to the original graph measure, is in many cases closer to the median of that measure than to its mean (especially, if its original distribution is heavy-tailed, e.g.\ power-law-like). This is due to the non-linearity of the transformation used in the presented measure standardisation procedure.\\ 
Total heights (positive/negative) on the right (G1, G2, G3) sum up to heights on the left (G2, G3, G4), respectively. That is, for a given year, subtracting the negative part from the positive part is the same operation for each column (generation). \\
Looking at the particular case at hand, we can immediately draw two conclusions: First, higher generation scores may result from destructive interference of lower generation measures: Additional information is gained/conserved. Second, taking specifically into account that China has been the most populous country during the whole observation period \cite{UN_population}, and assuming that migration is at least weakly correlated with the population size, China had a dramatically low centrality in the WMW relative to its size. This becomes clearer when looking at its abstract IN- and OUT-scores. The situation has changed after 1990, just like in the WTW; however, this change is more contained in the WMW. \\
One possible usage of the NGFP is to analyse in detail a node's (or even a group of nodes') centrality in a network by monitoring several scores, including abstract intermediate measures and their evolution over time. Note that a measure's score is always reflected by one or several physical real-world entities. After identifying such real-word contributions to a network measure, the NGFP can be used to quantify the influence of these contributions on a node's centrality. Considering the two cases at hand, such real-world contributions may be due to economic and immigration policies and their consequences. Given a particular single or composite score, this information can be of great use when, for example, directing futures policy decisions.\\
Furthermore, the universality of SM allows to compare scores for different networks and also to investigate their relations as it might be done for the WTW and the WMW. The NGFP offers a very useful tool to do so under a standard framework, which demonstrates once again the normative power of the presented methodology.
\section{Discussions}
\label{sec:IV}
\subsection{Validity}
\label{sec:IV-A}
In \hyperref[sec:II-B]{Section\ 2.2}, we mentioned that, given a \textit{representative data set}, we can perform a statistical measure normalisation in terms of the sample standard deviation $\sigma_s$, so that scores for different time instances are comparable to each other. By representative we mean that $\sigma_s$ is independent of the sample size. This is important since we want to express our results independently of the network size and only in terms of statistical deviations. Looking at the changes of $\sigma_s$ for all SM and all generations, it is clear that its correlation with the network size is weak as compared to correlations among different measures. From this, we conclude that the assumption of a representative data set is justified for both, the WTW and the WMW, while changes in the $\sigma_s$ are associated with underlying topological variation (evolution).\\
In \hyperref[sec:III-B]{Section\ 3.2}, we argued furthermore that, if the assumptions of the Lyapunov theorem \cite{Lyap_thm,prob_book1} hold, the CLT can be extended to non-identically distributed random variables, at least approximately. These assumptions are:
\begin{itemize}
\item \textit{Independence:} The $\bar{M}_i$ are independent of each other.
\item \textit{Lyapunov condition:} $\bar{M}_i$ are a sequence of, not necessarily identical, non-degenerate random variables (measures), with partial sum $S_n=\sum_{i=1}^{n}\,\bar{M}_i$ and variance $V_n=\sum_{i=1}^{n}\,\sigma_i^2=Var\left(S_n\right)$, and there exists a $\delta>0$ such that
\begin{equation}
\label{eq:lya}
\lim_{n\rightarrow\infty}\,\frac{\sum_{k}^{n}\,E\left(|\bar{M}_i|^{2+\delta}\right)}{V_n^{1+\delta/2}}=0,
\end{equation}
i.e.\ the $2+\delta\,$-moment exists.
\end{itemize}
Generally, it cannot be assured that the $2+\delta\,$-moment exists. Furthermore, we neither specify nor investigate a particular distribution of individual measures. Because of this inherent uncertainty, we choose the way of backward-engineering: Presuming that the necessary assumptions approximately hold, we use an \textit{a posteriori} hypothesis test, for each case at hand, to decide if the universal distribution (\ref{eq:logn}) is an acceptable description of the observed data. This is exactly what we have done by considering the outcomes of a KS-test in combination with a rather strict decision rule, whereby the resulting GoF is a measure for modelling accuracy in the particular case (see \hyperref[sec:III-B]{Section\ 3.2}). In fact, many low-generation SM considered in the SF show already an approximate normal distribution, which indicates the existence of higher moments, due to the smoothing of overall variability through the implemented measure standardisation procedure.\\
The requirement of \textit{independence} is generally hard to fulfil because many network measures are somehow mutually dependent, and thus correlated. By imposing an edge threshold in the cases at hand, the strength of a node is automatically coupled to its degree\footnote{The same is true for the maximal flow.}. Nevertheless, we might assume that those correlations are weakened for the case of SM, because each measure is transformed with an individual exponent $\lambda$. This is seen by the fast convergence of SM to the standard normal distribution, in contrary to the original measures (after solely statistical normalisation). In this case, one may talk about \textit{quasi-independence}, which is ultimately justified by the applicability of the approach.\\
An alternative approach would be to implement a least dependent component analysis \cite{least_dependent_component}. This is expected to reduce the dimensionality of the data set, while a drawback is the loss of interpredability of results, because individual measures can no longer be associated with particular flow processes on the network.\\
\subsection{Limitations}
\label{sec:IV-B}
\begin{figure}[!t]
\centering
\includegraphics[width=3.in]{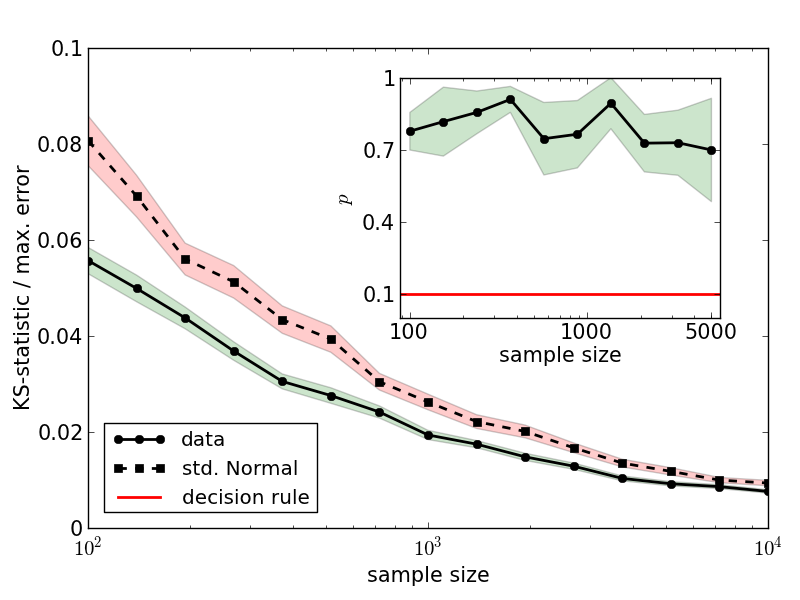}
\caption{(colour online) Inner panel: GoF of the SDH using the measure set $M_{\mathrm{arb}}$. The decision rule (DR) is given by the red horizontal line. Outer panel: Evolution of the KS-statistic of the standard normal distribution (\ref{eq:logn}) for samples drawn from $C_{\mathrm{comp}}^{\mathrm{arb}}$ (solid line) and the standard normal distribution itself (dashed line). Both panels show the evolution of average values with respect to the sample size $N$ including the $95\,\%$ confidence intervals (coloured areas) for $10$ (inner panel) and $100$ (outer panel) realizations, respectively.}
\label{fig:gof_ana}
\end{figure}

Considering the finiteness of $M_{\mathrm{SF}}$ together with the uncertainties about the validity of the underlying assumptions (see the \hyperref[sec:IV-A]{previous section}), the description of composite centralities by the standard normal distribution (null hypothesis) will surely not always be an acceptable description. Therefore, we define the decision rule (DR) $p\,>\,0.1$ (GoF from a KS-test, see \hyperref[sec:III-B]{Section\ 3.2}) whether to accept or reject the hypothesis.\\ 
The standard normal distribution of $C_{\mathrm{comp}}$ results from the interference between the single SM induced by the sampling. There are four major factors influencing the distribution of $C_{\mathrm{comp}}$: The number of measures $n$, the choice of measures $M$, the edge threshold $e_{\mathrm{th}}$ and the number of nodes $N$ (sample size), where the latter two items mainly concern the network at hand.\\
\begin{itemize}
\item $n:$ It is expected that the distribution of $C_{\mathrm{comp}}$ convergences to the distribution (\ref{eq:logn}) with an increasing number of sampled measures. Surprisingly, nearly all of the standardised measures in $M_{\mathrm{SF}}$ already pass the KS-test for both, the WTW and the WMW. Starting from G2, all sampled measures do so. The G1-measures where the KS-test for normality fails (mostly ASPL in both directions) are found to show (weak) bi-modal frequency distributions. Since uni-modality of raw measures is one of the requirements leading to (approximate) normality of the standardised measures, poor GoF values are expected in those cases. The rapid convergence observed for higher-generation measures is explained by ``destructive interference'' effects, cancelling multi-modality. 
\item $M:$ One expects that the GoF changes with the composition of $M$, since different measures generally follow different distributions and take different numerical values over various ranges. This is indeed the case. Given $M_{\mathrm{SF}}$, in order to improve the GoF, one might replace the G1-measure which generates the lowest single $p$-value by a measure which generates a higher value so as to improve the GoF. A single measure which generates a relatively high individual $p$-value, especially for the WTW, is the eigenvector centrality.\\
The results for the GoF using this alternative measure sets $M_{\mathrm{alt}}$\footnote{Note that the physical interpretation in terms of the D-R-T classification is now partly lost.} is shown in \hyperref[tab:p-vals]{Tab.\ 4}, where we also considered different edge thresholds (see below). The generally different reactions of the GoF for the WTW and the WMW to the use of $M_{\mathrm{alt}}$ instead of $M_{\mathrm{SF}}$ illustrates what might ultimately be attributed to the \textit{complexity} of the underlying systems: One  sees that just using ``well-behaved'' G1-measures does not guarantee a higher GoF.\\
%
%
\item $e_{th:}$ Neither a universal threshold dependence of $p$-values for the WTW or for the WMW is observed when looking at different value of $e_{th}$, nor particular ``good'' or ``bad'' values of the edge threshold can be identified. We conclude that the choice of the edge threshold does not have a particular effect on the GoF of the SNH for the description of CC-scores.
\item $N:$ Different instances of the WTW and the WMW have varying number of nodes. As can be seen in \hyperref[tab:p-vals]{Tab.\ 4}, the number of nodes does also not have a particular impact on the GoF (larger data sets are considered below).
\end{itemize}
One can see that our null hypothesis (\ref{eq:logn}) for the description of composite node centralities in the WTW and the WMW using the standard set $M_{\mathrm{SF}}$, as well as $M_{\mathrm{alt}}$, can be accepted in \textit{all} cases, and even for most single measures. To further test the SNH, we use the more stringent Shapiro-Wilk and Anderson-Darling tests for normality with the above-defined decision rule $p>.1$ or a $10\,\%$ significance level, respectively. Again, the SNH is not rejected in the great majority of cases given in \hyperref[tab:p-vals]{Tab.\ 3}. These results strongly support the claim of applicability of the proposed CC-framework for the universal description of node centralities across different evolving networks.\\
%
%
Both, the WTW as well as the WMW, are relatively small networks by their numbers of nodes compared to other real-world networks, such as the internet \cite{NW_book1,NW_book2}.
On the one hand, we are interested in how a composite centrality score behaves for different sample sizes (numbers of nodes or edges). On the other hand, we want to investigate the universality claim of the proposed framework to deal with \textit{arbitrary} sets of realistic measures, not necessarily originating from complex networks.\\
To achieve this, we consider a set $M_{\mathrm{arb}}$ of five ``measures'' whose values are drawn from five different continuous probability distributions stretching over hugely different numerical ranges, for which the particular choice has been rather arbitrary. The distributions and the values of their generic parameters are given in \hyperref[tab:p-vals]{Tab.\ 4}. Note that all measures fulfil the requirements of the Lyapunov theorem (see the \hyperref[sec:IV-A]{previous section}). For sample size $N=10^2-10^4$, we standardise these measures, compute the corresponding composite centrality scores $C_{\mathrm{comp}}^{\mathrm{arb}}$ and test the SDH, calculating the GoF by means of KS-tests\footnote{The KS-test is limited to $N=5000$ due to computational limitations.}. The results for the $p$-value (GoF) are shown in \hyperref[fig:gof_ana]{Fig.\ 5} (inner panel). It shows a slight degrading of the GoF as the sample size $N$ increases at the $95\,\%$ confidence level. This decrease in the GoF is expected and can be explained by the \textit{law of large numbers} from statistics (LLN) \cite{prob_book1,Lyap_thm}, which, in our case, can be written as 
\begin{equation}
\label{eq:lln}
P\left(\lim_{N\rightarrow\infty}\max\,\| \mathrm{CDF}_{\mathrm{SN}}^{\mathrm{distr.}}\,-\,\mathrm{CDF}_{\mathrm{SN}}^{\mathrm{sample\left(N\right)}}\|=0\right)=1\,.
\end{equation}
It states that the KS-statistic for synthetic samples drawn from (\ref{eq:logn}) converges to zero as the sample size increases (see \hyperref[sec:III-B]{Section\ 3.2}). On the other hand, the KS-statistic for data generated by sampling \textit{finitely} many random variables is expected to show a finite off-set (error) for increasing sample sizes, which illustrates the limits of our approximation for the description of $C_{\mathrm{comp}}$ using the standard normal distribution (\ref{eq:logn}).\\
However, for large samples, this off-set from the normal CDF is nothing else than the \textit{maximal error} when estimating the upper or lower bound probability for a certain CC-value of a randomly chosen node or edge\footnote{The maximal error for an interval is twice that value}. Thus, the universal scale (\ref{eq:logn}) can be used for an effective description of the distribution of composite centrality measures, taking into account a finite, supposedly small, error. This error depends on the particular set-up, like the number of measures and their distributions. Consequently, these distributions should be known at least approximately for a reliable and scalable error estimation.\\
\hyperref[fig:gof_ana]{Fig.\ 5} (outer panel) shows such an error estimation for $M_{\mathrm{arb}}$ and increasing sample size at the $95\,\%$ confidence level. As expected, the KS-statistic of synthetic samples undercuts the KS-statistic of $M_{\mathrm{arb}}$ at a certain point, which causes the decrease of GoF. The KS-statistic of $M_{\mathrm{arb}}$ is expected to level off with increasing sample size, as approximately displayed in the outer panel of \hyperref[fig:gof_ana]{Fig.\ 5}. The upper bound of the corresponding confidence interval is then interpreted as the maximal error when estimating the upper or lower bound probability for a certain CC-value at the given confidence level. Let us presume that the results given in \hyperref[fig:gof_ana]{Fig.\ 5} are obtained from an evolving network, e.g.\ at different time instances, and that $M_{\mathrm{arb}}$ represents some set of node measures. Considering an instance with approximately $N=10^4$ nodes, one can then use the outer panel of \hyperref[fig:gof_ana]{Fig.\ 5} to tell that a randomly chosen node will (or will not) have a composite centrality smaller or larger than a specified value, say the distribution's zero mean, with a certain probability subject to an error smaller than $1\,\%$ at the $95\,\%$ confidence level. The maximal errors is most likely attained in the middle part of the CDF (see \hyperref[fig:fit]{Fig.\ 2}). The real error for the extremes of the CC-spectrum, which might be of greater interest, can be assumed to be much smaller than that. This demonstrates clearly how the CC-framework can be applied, even in cases with low GoF for the SNH, providing one with extraordinary predictive power for the description of general complex systems. 
\section{Conclusions}
\label{sec:V}
Motivated by the observation of the lack of a unified framework to describe evolving complex systems in general and complex networks in particular, we have presented the concept of \textit{composite centrality} $C_{\mathrm{comp}}$ (CC-framework) to evaluate node and edge centralities based on a set of several graph measures. A standardisation procedure based on a non-linear transformation and statistical normalisation allows for a uniform comparison of different graph measures and their combinations over time and even across different networks. Based on a generalisation of the central limit theorem (GCLT), it is possible to derive a \textit{unified} and \textit{parameter-free} distribution function for the approximate description of $C_{\mathrm{comp}}$, which is given by the standard normal distribution (\ref{eq:logn}).\\
The implementation of invariant statistical inheritance schemes, leading from an initial set of measures to $C_{\mathrm{comp}}$, allows for the introduction of a variety of abstract centrality measures, which provide additional information. This is well depicted by the graphical analysis tool, which we label the \textit{network genetic fingerprint} (NGFP). As has been shown, the NGFP can be used when analysing a node's centrality and its evolution over time considering multiple measures simultaneously.\\
To test the proposed framework and its implications, we have considered the world trade web (WTW) and the world migration web (WMW), both on the weighted and directed level during a time period of 40 years. In both cases, we have concentrated on node measures, considering the largest strongly-connected component of a threshold graph. We have presented a standard framework (SF), which we proposed as a future para-\\digm for the analysis of general evolving complex networks, based on radial influence measures quantifying a node's centrality according to the criteria in terms of direction, range and texture. We have demonstrated how $C^{\mathrm{SF}}_{\mathrm{comp}}$ describes the evolution of node centralities and how it allows to compare scores for different networks using the unified scale given by the standard normal distribution (null hypothesis). \\
Since it is not possible to show \textit{a priori} that the assumptions of the GCLT hold for the general case, we have used an \textit{a posteriori} hypothesis test calculating a goodness-of-fit parameter to decide in what situations the unified description of node centralities through $C_{\mathrm{comp}}$ is appropriate. It turned out that the standard normal hypothesis is generally acceptable for a wide range of measure sets and network thresholds, making it a very useful tool for the analysis of general evolving complex networks.\\
To extend the CC-framework to general measure sets, not necessarily originating from complex networks, and to investigate the limitations of our null hypothesis, we have considered a set of measures drawn from five different distributions and largely varying sample sizes. We observed a small decrease of the goodness-of-fit with increasing sample sizes, which can be explained by the law of large numbers from statistics, and is thus understandable. Nevertheless, taking into account the relatively small error made by our approximation, the standard normal distribution can still be used for an \textit{effective} description of general composite centrality scores.\\
This universality is what may be called a \textit{natural scale} for complex networks. As we have shown, the proposed framework is not only limited to the study of evolving complex networks, but may also be used for the analysis of more general complex systems. Potential applications include the well-defined analysis of correlations and regressions between different metrics, which are widely used in all kinds sciences today.
\begin{table}[!ht]
\renewcommand{\arraystretch}{1.5}
\centering
\begin{tabular}{r|c|r|ll}
\multicolumn{5}{c}{WTW}\\
\hline
\hline
\multicolumn{1}{l|}{year} & \multicolumn{1}{c|}{N} & \multicolumn{1}{l|}{$e_{th}/10^6$} 
& \multicolumn{1}{l}{$p_{SF}$} & \multicolumn{1}{l}{$p_{\mathrm{alt}}$} \\
\hline
 & $184$ & $5$ & $.95$ & $.73$  \\
$1970$ & $181$ & $10$ & $.72$ & $.97$ \\
 & $173$ & $20$ & $.81$ & $.79$  \\
\hline
& $189$  & $5$ & $.80$ & $.53$  \\
$1980$ & $187$ & $10$ & $.53$ & $.45$  \\
& $177$  & $20$ & $.72$ & $.59$  \\
\hline
& $185$  & $5$ & $.83$ & $.52$  \\
$1990$ & $179$ & $10$ & $.64$ & $.78$  \\
& $173$  & $20$ & $.79$ & $.33$  \\
\hline
& $214$  & $5$ & $.94$ & $.98$  \\
$2000$ & $207$ & $10$ & $.95$ & $.90$  \\
& $195$  & $20$ & $.90$ & $.84$  \\
\hline
& $214$  & $5$ & $.97$ & $.81$  \\
$2010$ & $204$ & $10$ & $.92$ & $.88$ \\
& $195$  & $20$ & $.82$ & $.95$ \\
\multicolumn{5}{c}{}\\
\multicolumn{5}{c}{WMW}\\
\hline
\hline
\multicolumn{1}{l|}{year} & \multicolumn{1}{c|}{N} & \multicolumn{1}{l|}{$e_{th}/10^3$} 
& \multicolumn{1}{l}{$p_{SF}$} & \multicolumn{1}{l}{$p_{\mathrm{alt}}$} \\
\hline
  & $193$ & $1$ & $.99$ & $.76$ \\
$1960$ & $178$ & $2$ & $.99$ & $.89$ \\
  & $163$ & $4$ & $.83$ & $.90$ \\
\hline
 & $190$ & $1$ & $.98$ & $.71$ \\
$1970$ & $179$ & $2$ & $.97$ & $.98$ \\
 & $163$ & $4$ & $.90$ & $.70$ \\
\hline
 & $198$ & $1$ & $.82$ & $.25$ \\
$1980$ & $185$  & $2$ & $.66$ & $.92$ \\
 & $168$ & $4$ & $.91$ & $.72$ \\
\hline
 & $199$ & $1$ & $.56$ & $.84$ \\
$1990$ & $182$ & $2$ & $.94$ & $.98$ \\
 & $163$ & $4$ & $.91$ & $.82$ \\
\hline
 & $207$ & $1$ & $.69$ & $.74$ \\
$2000$ & $195$ & $2$ & $.85$ & $.99$ \\
 & $173$ & $4$ & $.96$ & $.88$\\
 \multicolumn{5}{c}{}
\end{tabular}
\caption{Results for the goodness-of-fit (GoF) KS-tests of the standard normal hypothesis (SNH) for the distribution of $C_{\mathrm{comp}}^{WTW}$ (upper table) and $C_{\mathrm{comp}}^{WMW}$ (lower table) during the corresponding 40-year periods. $p$-values are given for two different sets of first generation (G1) measures $M_{\mathrm{SF}}$ and $M_{\mathrm{alt}}$ and edge thresholds $e_{\mathrm{th}}$.}
\label{tab:p-vals}
\end{table}
%
%
%
\section*{Acknowledgment}
Special thanks to the developers and authors of \cite{python, ipython, nx, scipy}, which have been used for all numerical computations carried out in the present work. Thanks also given to the financial support of the Hong Kong Research Grants Council through the GRF Grant CityU 1109/12E. Additionally, we want to thank the reviewers for their many helpful comments on the original version of the manuscript.
%
%


\end{document}